\documentclass{article}
\usepackage{spconf,amsmath,graphicx,hyperref}

\usepackage{cite}
\usepackage{amsmath,amssymb,amsfonts}
\usepackage{algorithmic}
\usepackage{graphicx}
\usepackage{textcomp}
\def\BibTeX{{\rm B\kern-.05em{\sc i\kern-.025em b}\kern-.08em
    T\kern-.1667em\lower.7ex\hbox{E}\kern-.125emX}}

\usepackage{amsmath}
\usepackage{amssymb}
\usepackage{color}
\usepackage{xspace}
\usepackage{booktabs}
\usepackage{graphicx}

\usepackage{multirow}
\usepackage{tabularx}
\usepackage{tablefootnote}
\usepackage{booktabs}
\usepackage{makecell}
\usepackage{bbding}

\usepackage{amssymb}% http://ctan.org/pkg/amssymb
\usepackage{pifont}% http://ctan.org/pkg/pifont

\usepackage{microtype}
\usepackage{graphicx}
\usepackage{booktabs}
\usepackage{hyperref}
\usepackage{enumitem}
\usepackage{subfig}

\usepackage{booktabs}         % For professional-quality table rules
\usepackage[table]{xcolor}    % For cell and row colors
\usepackage{siunitx}          % For perfect decimal alignment of numbers
\usepackage{caption}          % For better caption formatting
\usepackage{etoolbox}         % For creating robust new commands

\newcommand{\todo}[1]{\textcolor{black}{#1}}

\newcommand{\tool}{\textsc{AnyAccomp}\xspace}

\title{\tool: Generalizable Accompaniment Generation via Quantized Melodic Bottleneck}
\name{Junan Zhang$^\star$ \qquad Yunjia Zhang$^\star$ \qquad Xueyao Zhang \qquad Zhizheng Wu\thanks{$^\star$Equal Contribution.}}
  
\address{The Chinese University of Hong Kong, Shenzhen}
\begin{document}
%\ninept
%
\maketitle
\begin{abstract}
Singing Accompaniment Generation (SAG) is the process of generating instrumental music for a given clean vocal input. However, existing SAG techniques use source-separated vocals as input and overfit to separation artifacts. This creates a critical train-test mismatch, leading to failure on clean, real-world vocal inputs. We introduce \tool, a framework that resolves this by decoupling accompaniment generation from source-dependent artifacts. \tool first employs a quantized melodic bottleneck, using a chromagram and a VQ-VAE to extract a discrete and timbre-invariant representation of the core melody. A subsequent flow-matching model then generates the accompaniment conditioned on these robust codes. Experiments show \tool achieves competitive performance on separated-vocal benchmarks while significantly outperforming baselines on generalization test sets of clean studio vocals and, notably, solo instrumental tracks. This demonstrates a qualitative leap in generalization, enabling robust accompaniment for instruments—a task where existing models completely fail—and paving the way for more versatile music co-creation tools. \todo{Demo audio and code: \url{https://anyaccomp.github.io/}}
\end{abstract}

\section{Introduction}
Singing Accompaniment Generation (SAG) is a crucial task in music creation, aiming to generate instrumental music that harmonically and rhythmically complements a given vocal melody~\cite{donahue2023singsong, chen2024fastsag}. High-fidelity SAG models hold immense potential to revolutionize the music creation process, offering powerful co-creation tools for artists, enabling producers to rapidly prototype ideas, and empowering amateurs to bring their musical visions to life.
% \todo{reference human and ai music co-creation}

\begin{figure}
    \centering
    \includegraphics[width=0.8\linewidth]{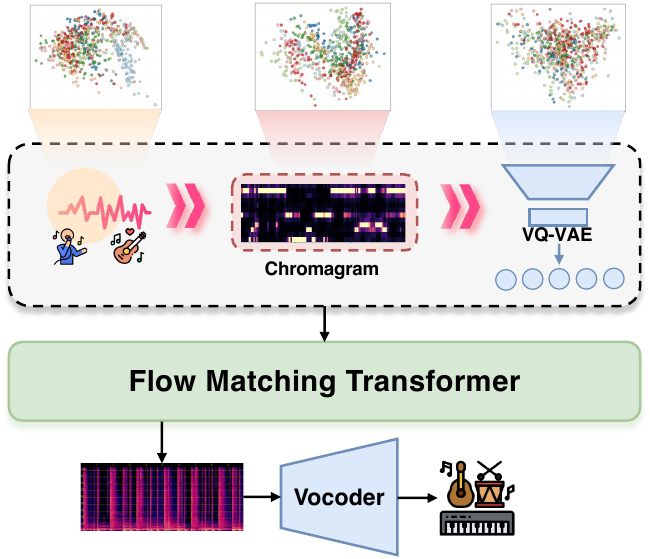}
    \caption{An overview of \tool. The process involves two main stages: (1) The input audio is processed through a quantized melodic bottleneck, where a VQ-VAE encodes its chromagram into a sequence of discrete tokens. (2) A Flow Matching Transformer generates a mel-spectrogram conditioned on these discrete tokens, which is subsequently synthesized into the final accompaniment audio by a vocoder. }
    \label{fig:overview}
    \vspace{-0.6cm}
\end{figure}

A common training paradigm for state-of-the-art SAG models~\cite{gong2025ace, donahue2023singsong, chen2024fastsag} relies on vocal data extracted by Music Source Separation (MSS) algorithms~\cite{luo2023bsrnn, lu2024bsroformer}. The vocal conditioning signal is typically a direct acoustic representation, such as the mel spectrogram used in FastSAG~\cite{chen2024fastsag}, or features from powerful Self-Supervised Learning (SSL) models, an approach pioneered by SingSong~\cite{donahue2023singsong} with a custom w2v-BERT~\cite{chung2021w2v}. However, both mel spectrograms and general-purpose SSL features are designed to preserve rich acoustic details. Consequently, when trained on imperfectly separated data, they invariably learn an artifactual correlation, associating valid accompaniment generation with the presence of separation artifacts. While previous works~\cite{donahue2023singsong, chen2024fastsag} attempt to mitigate this by injecting white noise, this can introduce new, unrealistic artifacts. This leads to a critical \textbf{train-test mismatch}: in real-world scenarios where recordings are clean, these models falter because they lack the very ``cues'' they were trained to expect. This dependency on imperfect data significantly limits the robustness and practical applicability of current SAG systems.

In this paper, we view the SAG task as a conditional generation problem, arguing the mismatch stems from the conditional representation's sensitivity to source-specific characteristics. We therefore propose a representation designed to be robust by satisfying two key properties: \textbf{timbre invariance} to filter out source-dependent textures such as timbre or separation artifacts, while \textbf{melodic clusterability} preserves the essential melodic content in a compact and structured form. This provides a unified condition that resolves the train-test mismatch, enabling a model trained solely on vocal data to robustly generalize to clean vocals and even instrumental tracks.

To this end, we propose \tool, a novel two-stage framework for robust accompaniment generation. The first stage of \tool is a bottleneck module that extracts a generalized melodic representation by using a Vector-Quantized Variational Autoencoder (VQ-VAE)~\cite{van2017neural} to quantize the input audio's timbre-invariant chromagram into a sequence of discrete codes. This process effectively isolates the core melodic content, creating a representation that is robust to source-dependent artifacts. The efficacy of this representation is empirically demonstrated in Figure~\ref{fig:empirical_experiments}, which shows our codes achieving significantly clearer melodic clusterability and timbre invariance compared to traditional mel spectrograms.

In the second stage, a Flow-Matching Transformer~\cite{lipman2022flow} generates the accompaniment conditioned on these robust and generalized representations. By decoupling the generation process from artifact-preserving representations, \tool avoids the correlations with separation artifacts that limit existing methods. Our experiments validate this approach: \tool not only performs competitively on in-domain separated-vocal benchmarks but also significantly outperforms baselines on generalization test sets with clean studio vocals and, notably, solo instrumental tracks. This success demonstrates that our method effectively resolves the critical train-test mismatch. The main contributions of this work are as follows:

\begin{itemize}[itemsep=1pt,topsep=0pt,parsep=0pt]
    \item We introduce a quantized melodic representation via a chromagram and VQ-VAE bottleneck, which is timbre-invariant and robust to separation artifacts. 
    
    \item We propose \tool, a two-stage framework that decouples accompaniment generation from artifact-prone inputs via a Flow-Matching Transformer, and demonstrate its effectiveness in resolving the train-test mismatch by outperforming baselines on clean studio vocals and solo instrumental tracks.
    % \item 
\end{itemize}

\section{\tool Framework}

\subsection{Quantized Melodic Bottleneck}
\label{sec:bottleneck}

The core of our approach is to learn an intermediate representation that captures essential melodic content while discarding irrelevant timbral information and source separation artifacts. To obtain this generalized representation, we employ a Vector-Quantized Variational Autoencoder (VQ-VAE) that operates on a 50 Hz Chromagram input $\mathbf{x}$. The model's encoder maps the chromagram to a continuous latent representation $\mathbf{z}_e(\mathbf{x})$, which is then quantized to its nearest neighbor $\mathbf{e}_k$ from a learned codebook $\mathbf{E}$ via $k = \arg\min_j \|\mathbf{z}_e(\mathbf{x}) - \mathbf{e}_j\|_2$. A decoder then reconstructs the chromagram $\hat{\mathbf{x}}$ from the resulting quantized sequence $\mathbf{z}_q(\mathbf{x})$. The model is trained to minimize a joint loss function:
\begin{equation}
    \mathcal{L} = \|\mathbf{x} - \hat{\mathbf{x}}\|_2^2 + \beta \|\mathbf{z}_e(\mathbf{x}) - \mathbf{z}_q(\mathbf{x})\|_2^2
\end{equation}

\begin{figure}[htbp]
    \centering
    \small

    % ---- 第一行: Timbre Invariance ----
    % 直接放置图例，并用 vspace 负值来缩小与下方图表的间距
    \includegraphics[width=0.9\linewidth]{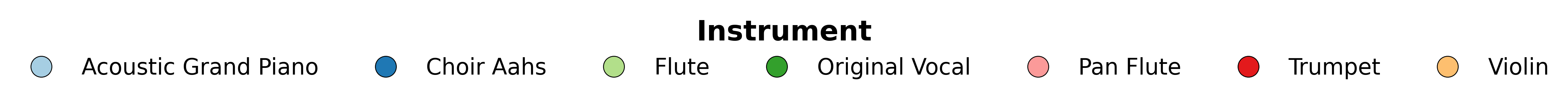}
    \vspace{-12pt} % <<--- 在这里减少垂直间距

    \subfloat[Mel\label{fig:timbre_mel}]{
        \includegraphics[width=0.24\linewidth]{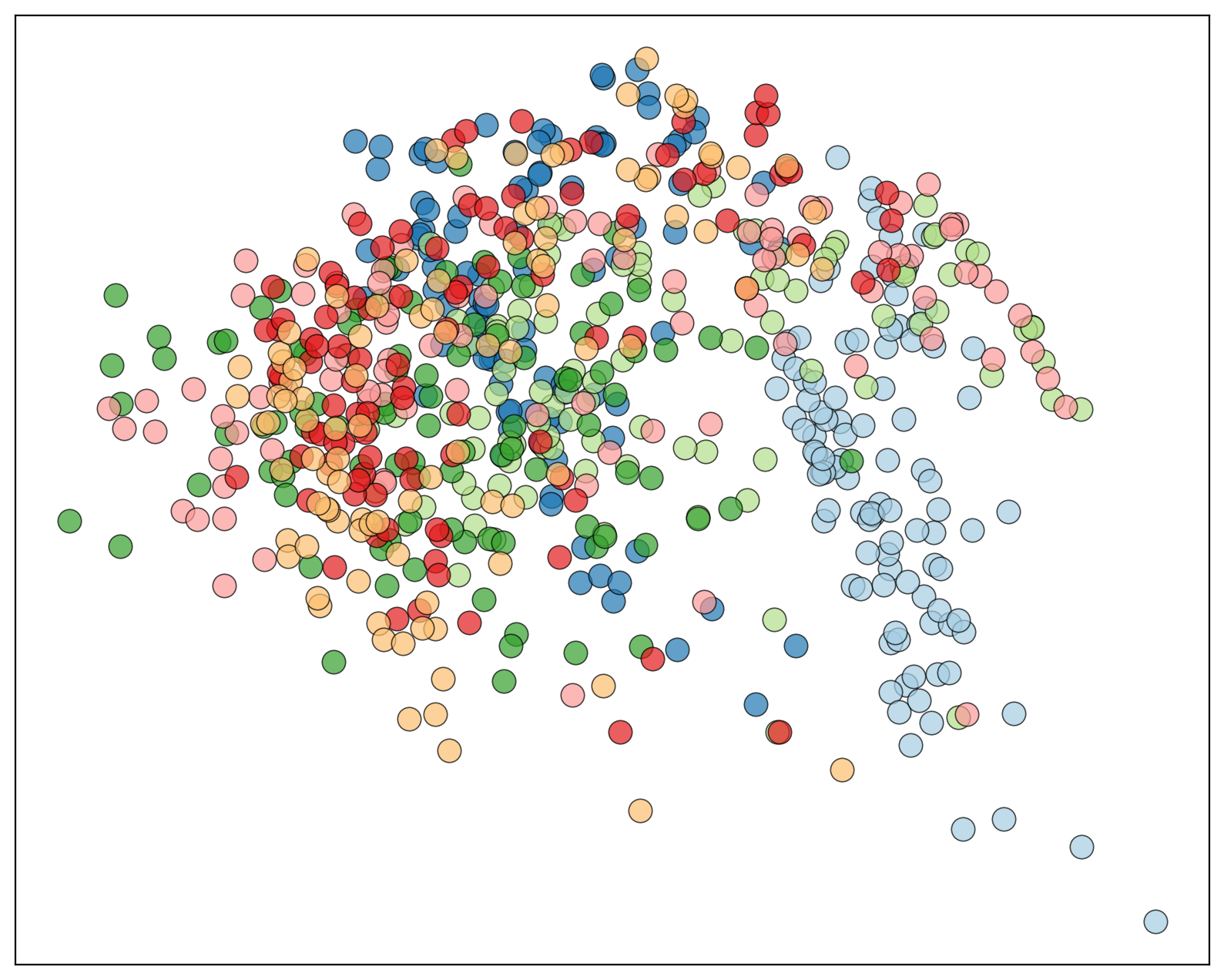}
    }
    \subfloat[MERT\label{fig:timbre_mert}]{
        \includegraphics[width=0.24\linewidth]{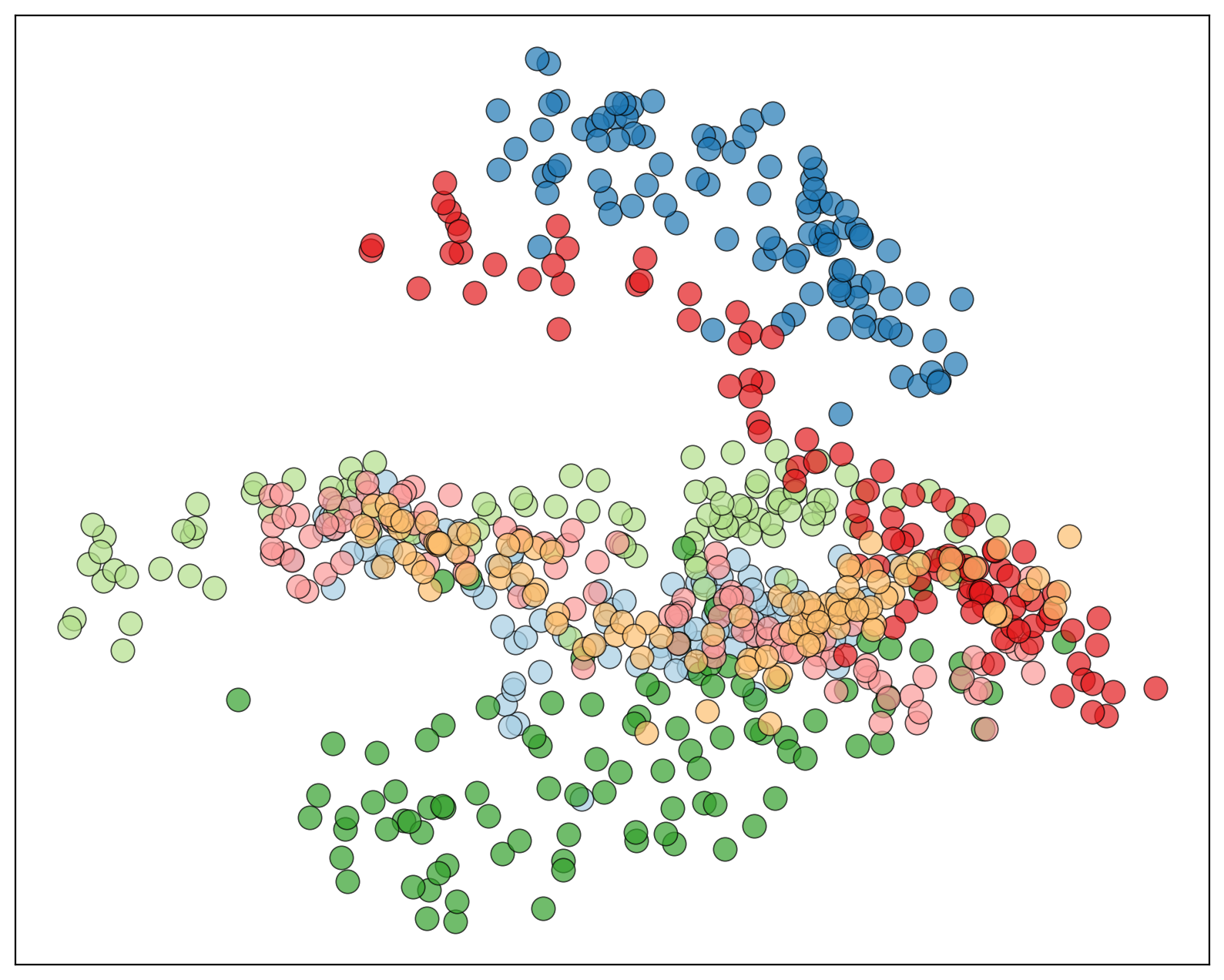}
    }
    \subfloat[Chroma (Dense)\label{fig:timbre_chroma_dense}]{
        \includegraphics[width=0.24\linewidth]{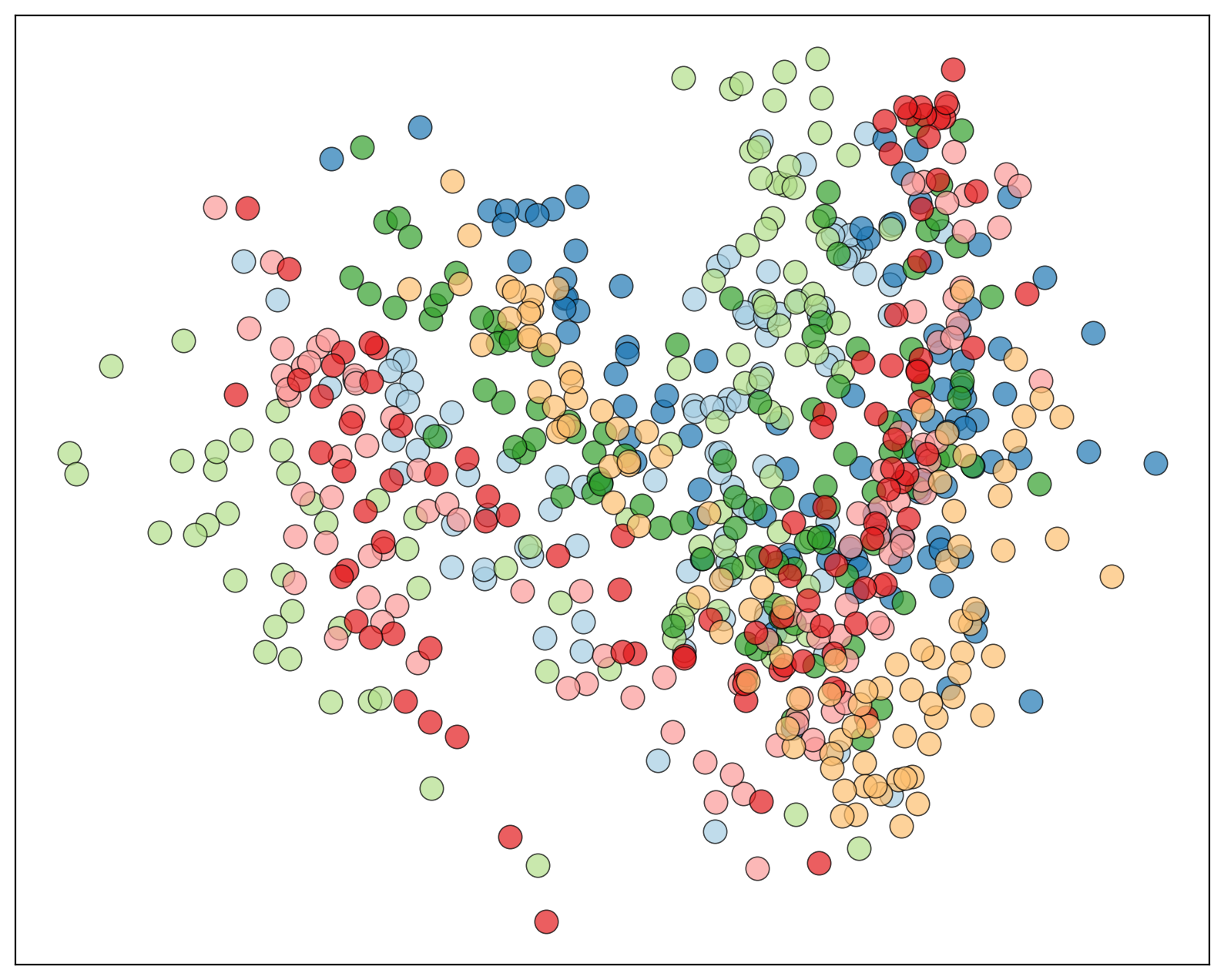}
    }
    \subfloat[Chroma (VQ)\label{fig:timbre_chroma_vq}]{
        \includegraphics[width=0.24\linewidth]{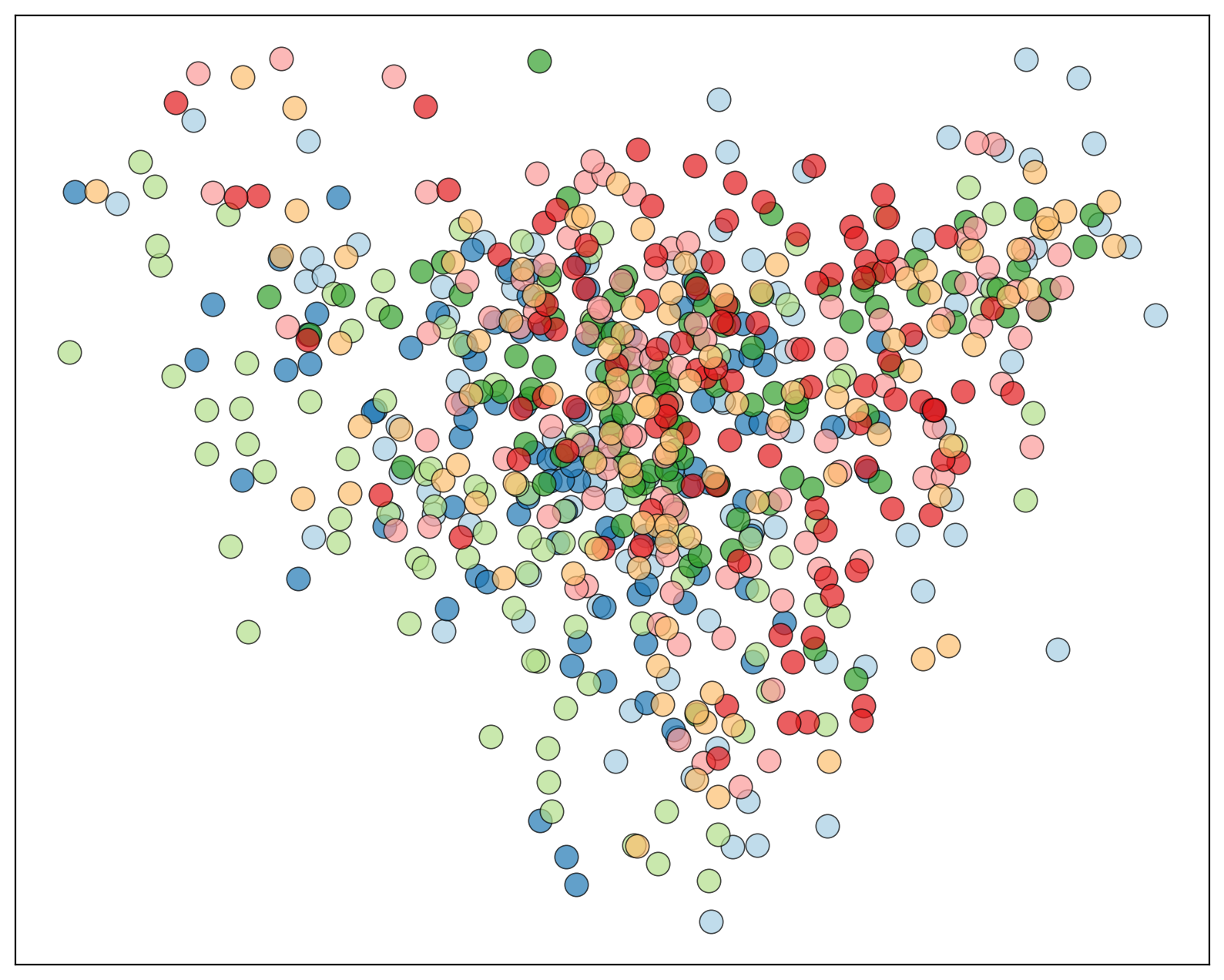}
    }
    
    % ---- 第二行: Melodic Clusterability ----
    % 同理，放置图例并减少下方间距
    \includegraphics[width=1\linewidth]{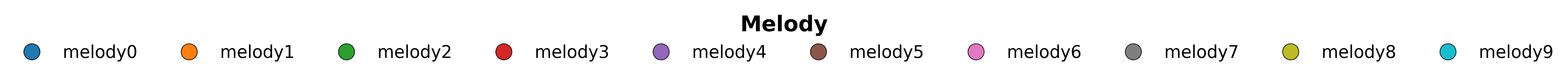}
    \vspace{-22pt} % <<--- 根据需要调整这个值，可以比第一个更大

    \subfloat[Mel\label{fig:melody_mel}]{
        \includegraphics[width=0.24\linewidth]{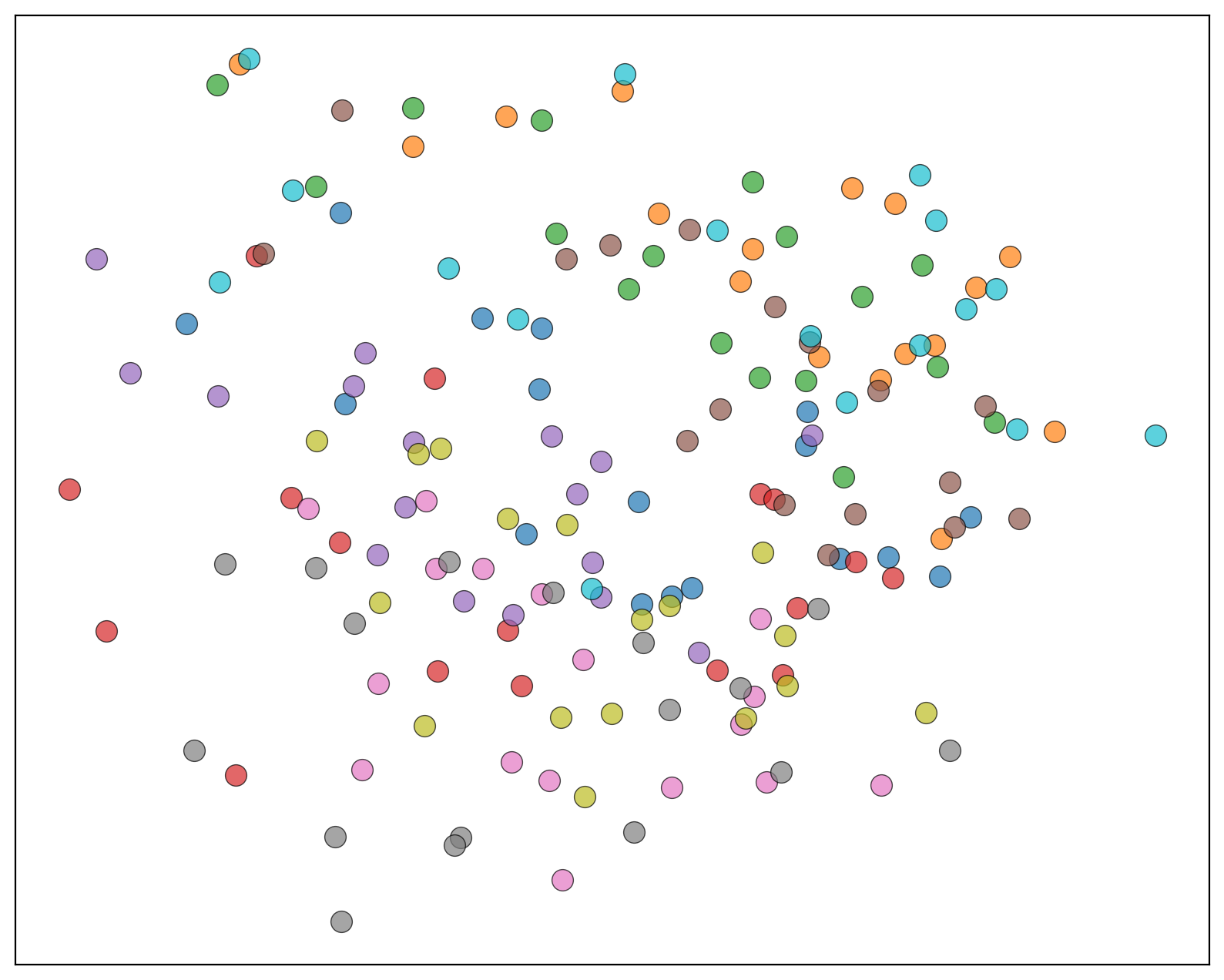}
    }
    \subfloat[MERT\label{fig:melody_mert}]{
        \includegraphics[width=0.24\linewidth]{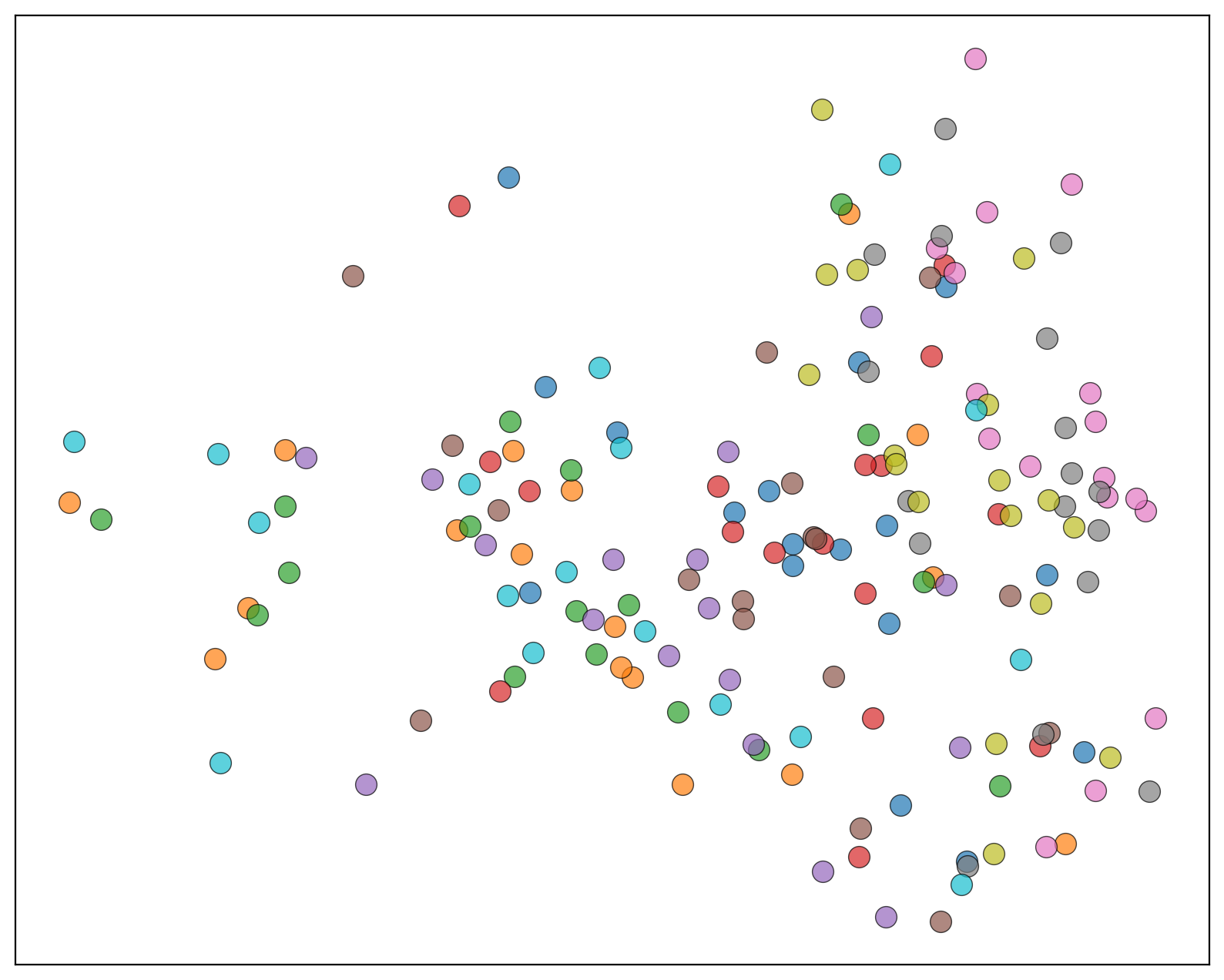}
    }
    \subfloat[Chroma (Dense)\label{fig:melody_chroma_dense}]{
        \includegraphics[width=0.24\linewidth]{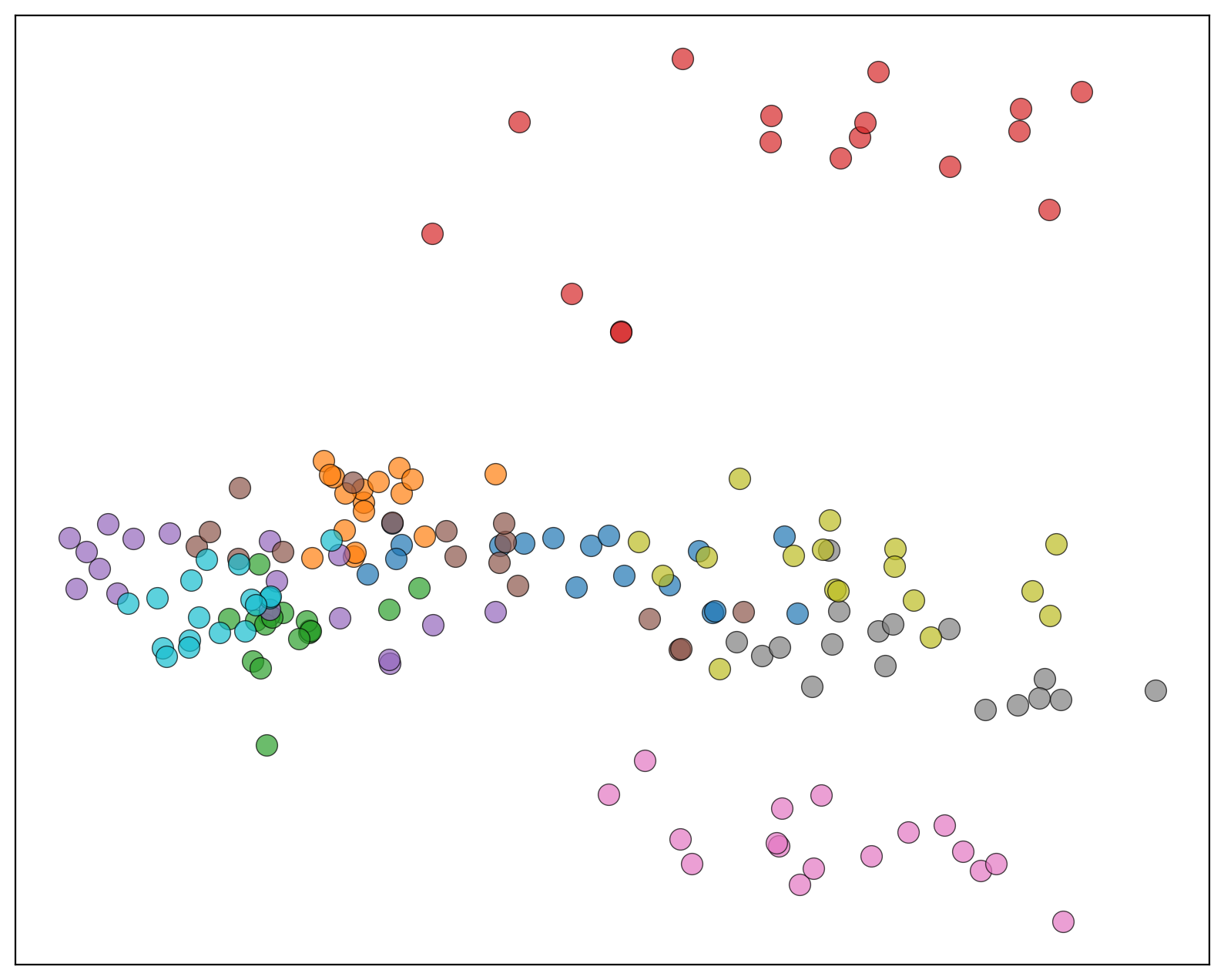}
    }
    \subfloat[Chroma (VQ)\label{fig:melody_chroma_vq}]{
        \includegraphics[width=0.24\linewidth]{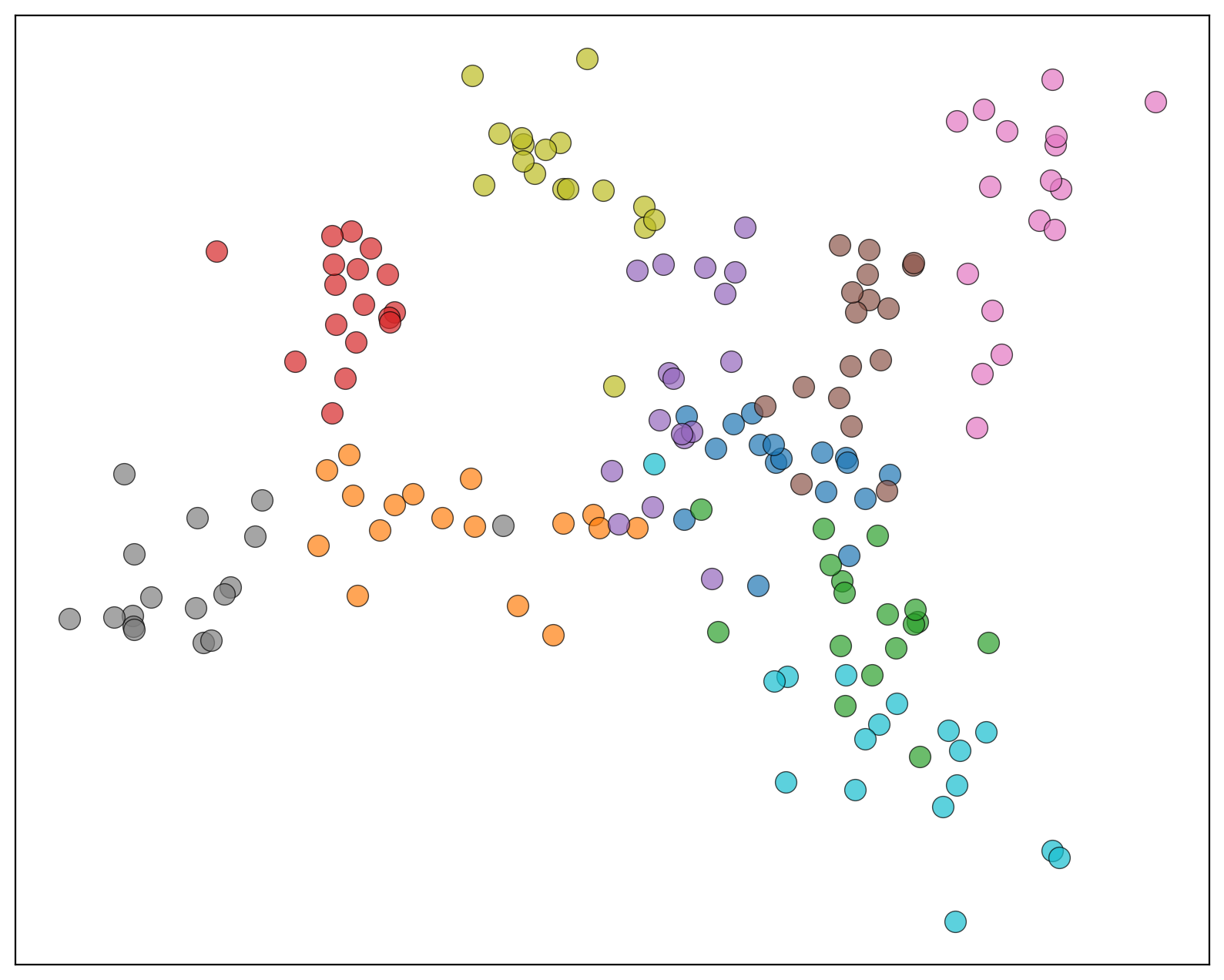}
    }
    
    \caption{
        Visualization of different representations. The top row evaluates \textbf{timbre invariance}, where better representations show instrument colors that are more thoroughly intermingled. The bottom row assesses \textbf{melodic clusterability}, where better representations form tighter, more distinct clusters for each melody color.
    }
    \label{fig:empirical_experiments}
    \vspace{-0.2cm}
\end{figure}

This quantization process forces the model to capture crucial melodic features while discarding artifacts, creating a robust bottleneck for the generation stage.

To evaluate our representation's effectiveness, we conduct a visualization experiment using melodies from the M4Singer dataset~\cite{zhang2022m4singer}. For each melody, we use the original vocal utterance and augment it by synthesizing several instrumental versions from its corresponding MIDI data using pretty\_midi\footnote{\url{https://github.com/craffel/pretty-midi}}. This creates a controlled set where the melody is constant while the timbre varies. From this combined collection, we extract and compare four representations: the standard mel spectrogram, features from the pre-trained MERT model~\cite{li2024mert}, a dense chromagram, and our VQ-quantized chromagram. Figure~\ref{fig:empirical_experiments} visualizes their PCA-reduced embeddings to evaluate two key properties.

% \textbf{Timbre invariance.} A robust representation must be agnostic to the source timbre, a property crucial for generalization. As visualized in the top row of Figure~\ref{fig:empirical_experiments}, the traditional mel spectrogram (Figure~\ref{fig:timbre_mel}) fails this test, exhibiting strong clustering by instrument type. This is particularly evident with the \texttt{Acoustic Grand Piano} (light blue), which forms a distinct group separate from other instruments. The dense chromagram (Figure~\ref{fig:timbre_chroma_dense}) significantly mitigates this bias by breaking up these obvious clusters. However, it is our final vector-quantized chromagram representation (Figure~\ref{fig:timbre_chroma_vq}) that achieves the ideal result: a feature space where points from all instruments are thoroughly blended, demonstrating a successful decoupling of melodic content from source timbre.

\textbf{Timbre invariance.} A robust representation must be agnostic to the source timbre, a property crucial for generalization. As visualized in the top row of Figure~\ref{fig:empirical_experiments}, representations that preserve rich acoustic features fail this test. The traditional mel spectrogram (Figure~\ref{fig:timbre_mel}) exhibits strong clustering by instrument type. This effect is even more pronounced with the MERT representation (Figure~\ref{fig:timbre_mert}), as its pre-training objective is designed to capture fine-grained acoustic characteristics. The dense chromagram (Figure~\ref{fig:timbre_chroma_dense}) significantly mitigates this bias by breaking up these obvious clusters. However, it is our final vector-quantized chromagram representation (Figure~\ref{fig:timbre_chroma_vq}) that achieves the ideal result: a feature space where points from all instruments are thoroughly blended, demonstrating a successful decoupling of melodic content from source timbre.

% \textbf{Melodic clusterability.} An effective representation must map identical melodies to compact regions in the feature space, irrespective of the instrument. The bottom row of Figure~\ref{fig:empirical_experiments} illustrates a clear progression. While the mel spectrogram (Figure~\ref{fig:melody_mel}) fails to form coherent clusters, the dense chromagram (Figure~\ref{fig:melody_chroma_dense}) shows a marked improvement, with distinct melodic clusters becoming visible. Our final vector-quantized chromagram representation (Figure~\ref{fig:melody_chroma_vq}) further refines this by producing exceptionally tight and well-separated clusters. This demonstrates that quantization not only preserves but also enhances the melodic structure, creating a highly robust and discrete condition for the generative model.

\textbf{Melodic clusterability.} An effective representation must map identical melodies to compact regions in the feature space, irrespective of the instrument. The bottom row of Figure~\ref{fig:empirical_experiments} illustrates a clear progression. Both the mel spectrogram (Figure~\ref{fig:melody_mel}) and the MERT representation (Figure~\ref{fig:melody_mert}) perform poorly, failing to form coherent melodic clusters. The dense chromagram (Figure~\ref{fig:melody_chroma_dense}) shows a marked improvement, with distinct melodic clusters becoming visible. Our final vector-quantized chromagram representation (Figure~\ref{fig:melody_chroma_vq}) further refines this by producing exceptionally tight and well-separated clusters, demonstrating that quantization not only preserves but also enhances the melodic structure, creating a highly robust and discrete condition for the generative model.

\subsection{Flow-Matching Accompaniment Generation}
% \todo{search for other paper?}
In the second stage, a Flow-Matching (FM) Transformer~\cite{lipman2022flow} generates the accompaniment conditioned on the discrete melodic codes. Crucially, this strategy decouples the generation process from input source artifacts, ensuring the output is based purely on the musical melody and rhythm.

We formulate accompaniment generation as a conditional flow-matching problem, where the goal is to learn a continuous-time probability path $p_t(\mathbf{x})$ that transforms a simple prior distribution into the target data distribution. Let $\mathbf{x}_1$ represent the target accompaniment mel spectrogram. The path begins from $\mathbf{x}_0 \sim \mathcal{N}(0, \mathbf{I})$, a sample from a standard Gaussian distribution. We define the trajectory between these two points as:
\begin{equation}
    \mathbf{x}_t = (1 - (1 - \sigma)t) \mathbf{x}_0 + t \mathbf{x}_1
\end{equation}
where $t \in [0, 1]$ is the timestep and $\sigma$ is a small constant (e.g., $10^{-5}$). During training, the timestep $t$ is sampled from a cosine schedule by first drawing $t' \sim U[0, 1]$ and then applying $t = 1 - \cos(t' \cdot \frac{\pi}{2})$ to concentrate sampling density near the beginning of the trajectory.

The model is trained to predict the vector field $\mathbf{v}_t$ that generates this path. The ground-truth velocity is the time derivative of $\mathbf{x}_t$:
\begin{equation}
    \mathbf{v}_t = \frac{d\mathbf{x}_t}{dt} = \mathbf{x}_1 - (1 - \sigma)\mathbf{x}_0
\end{equation}
Our model $f_\theta$, conditioned on the discrete melodic codes $\mathbf{c}$, learns to approximate this velocity field. The training objective is a simple mean squared error loss:
\begin{equation}
    \mathcal{L}_{FM}(\theta) = \mathbb{E}_{t, \mathbf{x}_1, \mathbf{x}_0, \mathbf{c}} \left\| f_\theta(\mathbf{x}_t, t, \mathbf{c}) - \mathbf{v}_t \right\|^2
\end{equation}

We add a Representation Alignment (REPA) loss~\cite{yu2024representation}, which aligns an intermediate FM Transformer layer with representations from a pre-trained music model~\cite{li2024mert}. The final objective is the weighted sum of the two losses: $\mathcal{L} = \mathcal{L}_{FM} + \lambda \mathcal{L}_{REPA}$.
% \begin{equation}
%     \mathcal{L} = \mathcal{L}_{FM} + \lambda \mathcal{L}_{REPA}
% \end{equation}

During inference, we generate the accompaniment mel spectrogram $\mathbf{x}_1$ by starting from a random noise sample $\mathbf{x}_0$ and integrating the predicted velocity $\hat{\mathbf{v}}_t = f_\theta(\mathbf{x}_t, t, \mathbf{c})$ from $t=0$ to $t=1$ using the forward Euler method: $\mathbf{x}_{t+h} = \mathbf{x}_t + h \hat{\mathbf{v}}_t$, where $h$ is the step size computed as the inverse of the total timesteps dedicated for integration. To leverage classifier-free guidance (CFG)~\cite{ho2022classifier}, we randomly drop the melodic condition $\mathbf{c}$ with a probability of 0.1 during the training process.

\section{Experiments}

\subsection{Experimental Setup}

\textbf{Training Data.}
We prepare 8k hours of paired singing voice–accompaniment data for training \tool, following the SingNet pipeline~\cite{gu2025singnet}. The data is sourced from in-the-wild songs on the Internet, separated into vocals and accompaniment, and then sliced into clips ranging from 3 s to 30 s. All audio is processed at a 24 kHz sampling rate.

\textbf{Implementation Details.} 
Our VQ-VAE (44M params), adapted from Amphion~\cite{amphion,amphion_v0.2}, quantizes a 24-bin chromagram into a 50 Hz sequence with a 512-entry codebook, and is trained for 0.5M steps with a batch size of 200 seconds. The Flow-Matching (FM) Transformer, based on Vevo~\cite{zhang2025vevo,zhang2025vevo2}, consists of 10 LLaMA decoder layers with a hidden dimension of 1024, totaling 220M parameters. We fine-tune the vocoder from Vevo on our music data for use in our system. FM models are trained for 1M steps with a per-GPU batch size of 100 seconds, incorporating REPA loss with MERT-330M~\cite{li2024mert} as the alignment target at layer 4 of $\lambda=0.5$ weight. All models are optimized with AdamW~\cite{adamw} (learning rate $1\text{e-}4$, 32k warmup steps) and trained on a single GPU. During inference, we use 50 sampling steps with a CFG scale of 3.

\textbf{Compared Methods.}
To evaluate our melodic bottleneck, we compare \tool with FastSAG~\cite{chen2024fastsag}, a state-of-the-art non-autoregressive model using vocal mel-spectrograms, and two controlled variants that share the FM Transformer architecture of \tool: FM-Mel, conditioned on mel spectrograms with added white noise (15-20 dB SNR), and FM-Chroma, conditioned on raw 24-bin chromagrams. \tool instead uses quantized chromagram codes.

\begin{table}[htpb]
    \centering
    \caption{Objective evaluation results showing that \tool is competitive on in-domain separated vocals (YuE) while strongly outperforming baselines on generalization to clean vocals (MUSDB18) and instruments (MoisesDB).}
    \setlength{\tabcolsep}{4pt}
    \label{tab:objective_grouped}
    \small
    \begin{tabular}{lccccccc}
    \toprule
    \textbf{Model} & \textbf{APA$\uparrow$} & \textbf{FAD$\downarrow$} & \textbf{CE$\uparrow$} & \textbf{CU$\uparrow$} & \textbf{PQ$\uparrow$} & \textbf{PC$-$} \\
    \midrule
    \rowcolor{gray!20}
    \multicolumn{7}{c}{\textbf{YuE (In-domain Separated Vocal)}} \\
    \midrule
    Ground Truth & - & - & 7.270 & 7.784 & 7.734 & 5.752 \\
    \cline{1-7}
    FastSAG & 0.444 & 0.598 & 6.351 & 6.821 & 6.814 & 6.321 \\
    FM-Mel & \textbf{0.806} & 0.416 & 6.964 & 7.725 & 7.758 & 5.614 \\
    FM-Chroma & 0.633 & 0.418 & 7.151 & 7.801 & 7.909 & 5.436 \\
    \tool & 0.713 & \textbf{0.414} & \textbf{7.283} & \textbf{7.903} & \textbf{7.989} & 5.742 \\
    \midrule
    \rowcolor{gray!20}
    \multicolumn{7}{c}{\textbf{MUSDB18 (Clean Vocal)}} \\
    \midrule
    Ground Truth & - & - & 7.164 & 7.616 & 7.485 & 5.957 \\
    \cline{1-7}
    FastSAG & 0.000 & 1.115 & 4.853 & 5.789 & 6.315 & 5.778 \\
    FM-Mel & 0.167 & 0.999 & 5.202 & 6.616 & 6.841 & 4.090 \\
    FM-Chroma & 0.704 & 0.798 & 7.017 & 7.598 & 7.744 & 5.104 \\
    \tool & \textbf{0.710} & \textbf{0.788} & \textbf{7.277} & \textbf{7.804} & \textbf{7.891} & 5.498 \\
    \midrule
    \rowcolor{gray!20}
    \multicolumn{7}{c}{\textbf{MoisesDB (Instruments)}} \\
    \midrule
    Ground Truth & - & - & 7.236 & 7.791 & 7.778 & 5.694 \\
    \cline{1-7}
    FastSAG & 0.000 & 0.904 & 5.966 & 6.507 & 6.696 & 5.952 \\
    FM-Mel & 0.000 & 0.936 & 5.424 & 6.923 & 7.151 & 3.804 \\
    FM-Chroma & 0.157 & \textbf{0.849} & 6.308 & 7.377 & 7.508 & 4.110 \\
    \tool & \textbf{0.203} & 0.890 & \textbf{6.660} & \textbf{7.581} & \textbf{7.581} & 4.798 \\
    \bottomrule
    \end{tabular}
    % \vspace{-0.3cm}
\end{table}

\textbf{Evaluation Datasets.} We evaluate our models on three distinct 10-second clip datasets to test different capabilities. For in-domain performance, we use 3,000 separated vocal-accompaniment pairs from the YuE data~\cite{yuan2025yue}. To assess generalization to artifact-free audio, we use 2,777 clean vocal stems from the MUSDB18 test set~\cite{rafii2017musdb18}. Finally, to push the boundaries of generalization, we evaluate on 2,500 solo instrumental tracks from MoisesDB~\cite{pereira2023moisesdb}.

\textbf{Evaluation Metrics.} We assess accompaniment with objective and subjective metrics. Objective measures include Fréchet Audio Distance (FAD)~\cite{kilgour2018fr} for distributional similarity, Accompaniment Prompt Adherence (APA)~\cite{grachten2025apa} for condition alignment, and audiobox-aesthetics~\cite{tjandra2025meta} scores on Content Enjoyment (CE), Content Usefulness (CU), Production Complexity (PC), and Production Quality (PQ). For subjective evaluation, we ran a MOS test on 20 samples per test set, where listeners rated overall quality and coherency on a 1–5 scale.

\subsection{Results}

The objective results in Table~\ref{tab:objective_grouped} confirm our bottleneck representation's pivotal role in robust generalization. While \tool is competitive on the in-domain YuE dataset, its true advantage emerges where the train-test mismatch becomes critical. On clean vocals (MUSDB18), mel-spectrogram-conditioned models (FastSAG, FM-Mel) suffer a catastrophic collapse, with APA scores drop to 0. This score signifies a complete failure in conditional generation, as the resulting accompaniment has no correlation to the input melody—a direct consequence of severe overfitting to separation artifacts. This failure is even more pronounced on the more out-of-domain instrumental tracks from MoisesDB.

In stark contrast, \tool, leveraging by its quantized melodic bottleneck, maintains high coherence and quality across all generalization sets. This demonstrates its superior generalization capability (validated by the timbre invariance and melodic clusterability shown in Figure~\ref{fig:empirical_experiments}), enabling it to overcome the train-test mismatch and robustly generate accompaniment for entirely new timbres where other models completely fail. We also note that Production Complexity (PC) measures textural complexity, not perceptual quality, and is thus not a ``higher-is-better'' metric.

\begin{figure}[ht]
    \centering
    \includegraphics[width=1\linewidth]{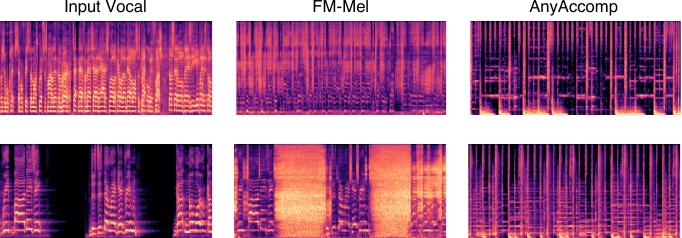}
    \vspace{-0.5cm}
    \caption{A case study on a clean MUSDB18 vocal. FM-Mel's output exhibits severe \textbf{spectral leakage} from the input, a sign of overfitting to source-separation artifacts. In contrast, \tool generates a coherent instrumental accompaniment.}
    \label{fig:casestudy}
    \vspace{-0.2cm}
\end{figure}

\begin{table}[htpb]
    \centering
    \caption{Subjective evaluation results on the three test sets. Quality refers to the overall accompaniment quality, while Coherency measures how well the accompaniment matches the input.}
    \small
    \label{tab:subjective_grouped}
    \begin{tabular}{l cc}
    \toprule
    \textbf{Model} & \textbf{Quality$\uparrow$} & \textbf{Coherency$\uparrow$} \\
    \midrule
    \rowcolor{gray!20}
    \multicolumn{3}{c}{\textbf{YuE (In-domain Separated Vocal)}} \\
    \midrule
    Ground Truth & 3.92 & 3.88 \\
    \cline{1-3}
    FastSAG      & 1.98 & 1.82 \\
    \tool & \textbf{3.12} & \textbf{3.05} \\
    \midrule
    \rowcolor{gray!20}
    \multicolumn{3}{c}{\textbf{MUSDB18 (Clean Vocal)}} \\
    \midrule
    Ground Truth & 3.65 & 3.48 \\
    \cline{1-3}
    FastSAG      & 1.73 & 1.48 \\
    \tool & \textbf{3.23} & \textbf{2.75} \\
    \midrule
    \rowcolor{gray!20}
    \multicolumn{3}{c}{\textbf{MoisesDB (Instruments)}} \\
    \midrule
    Ground Truth & 4.05 & 4.08 \\
    \cline{1-3}
    FastSAG      & 1.62 & 1.52 \\
    \tool & \textbf{3.00} & \textbf{2.70} \\
    \bottomrule
    \end{tabular}
\vspace{-0.3cm}
\end{table}

As illustrated in the case study of Figure~\ref{fig:casestudy}, FM-Mel suffers from severe spectral leakage, directly copying artifacts from the input vocal. In contrast, \tool generates a coherent accompaniment, highlighting its robustness against overfitting.

The subjective listening result (Table~\ref{tab:subjective_grouped}) confirms our objective findings. Listeners rated \tool significantly higher than FastSAG in both Quality and Coherency across all datasets. Crucially, \tool maintains high scores on the challenging MUSDB18 and MoisesDB generalization sets, validating that our objective improvements translate to a superior and more robust listening experience.

\section{Conclusion}
In this work, we present \tool, a framework that resolves the critical train-test mismatch in SAG models by using a quantized chromagram bottleneck to decouple the generation process from source-separation artifacts and timbre variations. Our model achieves competitive in-domain performance while drastically outperforming baselines when generalizing to clean vocals and unseen solo instruments. Future work will focus on optimizing bottleneck parameters (e.g., frame rate and vocabulary size), and exploring alternative representations like the Constant-Q Transform (CQT).

\ninept
\bibliographystyle{IEEEbib}
\bibliography{ref}

\end{document}